\documentclass[journal=tches, preprint]{iacrtrans}

\makeatletter
\def\ps@title{%
  \fancyhf{} 
  \renewcommand{\headrulewidth}{0pt}
  \renewcommand{\footrulewidth}{0pt}
}
\makeatother

\usepackage{multirow}
\usepackage{algorithm}
\usepackage{array}
\usepackage{makecell}
\usepackage{enumitem} 

\usepackage{algpseudocode}
\usepackage{booktabs}
\usepackage{listings}
\usepackage{xcolor}
\usepackage{amsmath, amssymb}

\usepackage{tikz}
\usetikzlibrary{positioning,arrows.meta,calc,fit,backgrounds,shapes.geometric,decorations.pathreplacing}

\usetikzlibrary{arrows}
\usetikzlibrary{calc}
\usetikzlibrary{positioning}

\tikzset{XOR/.style={draw,circle,append after command={
        [shorten >=\pgflinewidth, shorten <=\pgflinewidth,]
        (\tikzlastnode.north) edge (\tikzlastnode.south)
        (\tikzlastnode.east) edge (\tikzlastnode.west)
        }
    }
}

\lstdefinelanguage{armasm}{
  morekeywords={subs,cmp,cmn,eor,orr,orreq,and,andeq,ite,itt,mvn,mvneq,
                mov,moveq,movne,lsr,lsl,asr,rsb,sub,add,ldr,ldrd,str,
                push,pop,vpush,vpop,vmov,vldm,vstm,b,bl,bx},
  sensitive=true, comment=[l]{@}, morecomment=[l]{//},
}
\newcommand{\Sref}[1]{Section~\ref{#1}}
\newcommand{\Tref}[1]{Table~\ref{#1}}

\newcommand{\Aref}[1]{Algorithm~\ref{#1}}

\usetikzlibrary{decorations.pathreplacing}

\newcommand{\HQC}[1]{\textsf{HQC-#1}}

\usepackage[hyphens]{url}
\usepackage[colorlinks]{hyperref}
\author{Jihoon Jang\inst{1}, Hanbeom Shin\inst{1}, Suhri Kim\inst{2}, Seokhie Hong\inst{3}, \\Donggeun Kwon\inst{4}\thanks{Corresponding author.}}
\institute{
    Korea University, Seoul, South Korea, \email{{jhw1031506, newonetiger}@korea.ac.kr} 
    
    \and 
    Sungshin Women’s University, Seoul, South Korea, 
    \email{suhrikim@sungshin.ac.kr}

    \and 
    SmartM2M, Busan, South Korea, 
    \email{shhong@smartm2m.co.kr}

    \and 
    Kunsan National University, Gunsan, South Korea, 
    \email{dgkwon@kunsan.ac.kr}
}
\authorrunning{J. Jang et al.}

\title[Optimizing HQC on ARM Cortex-M4]{Optimizing Polynomial Multiplication and Fixed-Weight Sampling for HQC on ARM Cortex-M4}

\begin{document}

\maketitle


\begin{abstract}
In this paper, we present an optimized implementation of Hamming Quasi-Cyclic (HQC) on the ARM Cortex-M4. We optimize (i) the polynomial multiplication and (ii) the support expansion in fixed-weight sampling, and (iii) propose an optional caching strategy that reuses the public transforms and hash recomputed under a fixed key.
For the polynomial multiplication, the fixed-constant multiplications in the Frobenius additive FFT (FAFFT) butterfly spend nearly half of their instructions on VMOV data movements between general-purpose and floating-point registers rather than arithmetic.
Because minimizing the XOR count alone can increase the total instruction count, we propose a dirty-aware register-allocation policy and an XOR-operation reordering that reduce the VMOV count by up to $48.1\%$ while leaving the XOR count unchanged. We apply these to a multiplication that combines prior FAFFT-CRT methods, and for \HQC{1} we further find a 34\% sparser FAFFT modulus that lowers the CRT reconstruction cost.
For fixed-weight sampling, we rewrite the support expansion with predicated execution and 4-way unrolling, lowering the per-word cost of its inner loop from $22$ to $6$ cycles while remaining constant-time.

On the NUCLEO-L4R5ZI board, our implementation reduces key generation, encapsulation, and decapsulation by up to $33.1\%$, $34.6\%$, and $29.8\%$ over the faster of the two prior state-of-the-art implementations, and the optional caching yields a further reduction of up to $32.7\%$ and $18.9\%$ for encapsulation and decapsulation.
\end{abstract}

\section{Introduction}
\label{sec:intro}

In March 2025, NIST selected the code-based Hamming Quasi-Cyclic (HQC)~\cite{hqc-spec} scheme as an additional post-quantum KEM standard to provide algorithmic diversity beyond the lattice-based ML-KEM~\cite{nist-ir-8545}. Unlike ML-KEM, the previously selected PQC KEM, the security of HQC reduces to the syndrome decoding problem for quasi-cyclic codes, thereby providing diversity that does not rely on lattice-based assumptions. In \HQC{1}, \HQC{3}, and \HQC{5}, polynomial multiplication is performed over the rings $\mathbb{F}_2[x]/(x^n - 1)$ with $n = 17{,}669$, $35{,}851$, and $57{,}637$, respectively. Together with polynomial multiplication, sampling and hashing largely determine the overall performance of HQC, while decapsulation additionally incurs the cost of decoding.



Research on accelerating HQC on the Cortex-M4 has progressed actively around NIST's Round 4 selection, from the Frobenius additive FFT implementation of Lee et al.~\cite{MyeonghoonIOT} to the FAFFT-CRT methods that Jang et al.~\cite{jlk25} and Chen et al.~\cite{TCHES:CCC26} independently proposed to avoid zero-padding overhead, both reducing the transform size and the number of XOR operations. The actual instruction cost, however, is not determined solely by the XOR count. Since the Cortex-M4 provides only a limited number of GP (general-purpose) registers, data transfers between GP and VFP (vector floating-point) registers occur frequently, so an XOR sequence with fewer XOR operations may incur more VMOV instructions under register pressure and thus a larger total instruction count. In addition, fixed-weight sampling requires repeated mask generation when expanding a support into a dense bit vector, and reusing a fixed key across multiple sessions repeats the same polynomial generation and transform. To reduce these costs, we propose implementation-level optimizations tailored to the ARM Cortex-M4.

\subsection{Our Contributions}

The main contributions of this paper are summarized as follows:

\begin{itemize}
    \item \textbf{Optimizing HQC Polynomial Multiplication.} 
    We combine algorithm-level optimizations of FAFFT-CRT with register-level optimizations of bit-sliced butterfly assembly. First, we integrate truncated basis conversion and the sharing of the forward transform of the common operand $\mathbf{r}_2$ into the FAFFT-CRT implementation of Jang et al.~\cite{jlk25}. We also search for a new FAFFT evaluation set and reduce the Hamming weight of the FAFFT modulus polynomial for \HQC{1} from 165, as reported by Chen et al.~\cite{TCHES:CCC26}, to 109, a reduction of 34\%, thereby lowering the reconstruction cost. Next, we show that VMOV instructions account for nearly half of all instructions in bit-sliced butterfly assembly, and that minimizing only the XOR count could instead increase the total instruction cost. To address this issue, we propose a new dirty-aware register-allocation policy and an XOR-operation reordering method that uses the costs of load and spill VMOV instructions as its objective function. As a result, while preserving the EOR count of the butterfly operations, our method reduces the number of VMOV instructions over the state-of-the-art implementation by 40.6\%, 48.1\%, and 37.6\% for \HQC{1}, \HQC{3}, and \HQC{5}, respectively.

    \item \textbf{Optimizing Support Expansion in Fixed-Weight Sampling.} 
    We propose rewriting the inner loop of \texttt{vect\_write\_support\_to\_vector} (WSV), which expands the support generated by HQC sampling into a dense bit vector, using ARM predicated instructions. To avoid data-dependent timing, the baseline implementation distributed by the HQC team spends six ALU operations per index generating a full-width mask, which an AND and an OR then apply to the output word. However, generating and applying this mask takes a substantial portion of the inner-loop cost. We use the Cortex-M4 IT block and \texttt{ORREQ} instructions to directly perform the conditional OR operation, thereby allowing the implementation to remain constant-time without a mask. We further combine this approach with 4-way outer-loop unrolling. As a result, our method reduces the per-word cost of the inner loop from approximately 22 cycles to 6 cycles while remaining constant-time. WSV is invoked 9 times during a single KEM cycle, and approximately 1.2 million inner iterations are performed for \HQC{5}.

    \item \textbf{Improved HQC Performance on Cortex-M4.} Evaluating the implementation of the two methods above on a NUCLEO-L4R5ZI board, we reduce key generation, encapsulation, and decapsulation by $27.7$--$33.1\%$, $27.6$--$34.6\%$, and $22.0$--$29.8\%$ relative to the lower cycle count of the two prior state-of-the-art implementations, Chen et al.~\cite{TCHES:CCC26} and Jang et al.~\cite{jlk25}, for each operation. We further propose an optional caching technique for environments that handle many sessions under the same public key. It precomputes the regenerated polynomial, its FAFFT forward transform, and the public-key hash once and reuses the stored results instead of recomputing them on every encapsulation and decapsulation. In an experiment assuming that the same public key is reused, this technique reduces encapsulation and decapsulation by up to $32.7\%$ and $18.9\%$ compared to the case without caching.
\end{itemize}


\subsection{Related Work}
\label{sec:related}

Binary polynomial multiplication is a major performance bottleneck in HQC, and approaches for accelerating it fall broadly into divide-and-conquer methods and FFT-based methods. Implementations such as \texttt{gf2x}~\cite{gf2xlib} use schoolbook, Karatsuba, and Toom--Cook multiplication~\cite{brent2008faster, von1996arithmetic, camara2013fast}, with the final word-level products computed by carry-less multiplication, which propagates no carries. On platforms equipped with dedicated carry-less multiplication instructions, such as \texttt{PCLMULQDQ} on x86 and \texttt{PMULL} on Neon, Karatsuba-based methods have been reported to outperform FFT-based methods for operand sizes below approximately $20\,$kbit~\cite{ACNS:CGKT22}. The ARM Cortex-M4 targeted in this work, however, provides no such instruction, which makes FFT-based multiplication a competitive alternative.

In SIMD environments, Jang et al. combined branch-free Toom--Cook/Karatsuba decomposition with \texttt{VPCLMULQDQ}/\texttt{PMULL}-based multiplication, reducing the multiplication cycles of \HQC{1} and \HQC{3} by $17$--$32\%$ and observing that FFT-based methods may be more advantageous for \HQC{5}~\cite{jang2026faster, drucker2020fast}, while Chen et al. combined additive-FFT multiplication with the Chinese Remainder Theorem (CRT) and added a caching strategy for transformed key polynomials~\cite{TCHES:CCPY26}.

On the Cortex-M4, Lee et al. first applied FAFFT to HQC with bit-sliced butterflies, but since the polynomial lengths of HQC slightly exceed powers of two, the inputs had to be zero-padded to length $65{,}536$, incurring substantial transform overhead~\cite{MyeonghoonIOT}. A subsequent study~\cite{jlk25} addressed this padding overhead by lowering the transform size to $32{,}768$ through hybrid FAFFT--CRT and FAFFT--Karatsuba methods and, combining radix-16 multiplication with butterfly XOR-sequence reduction and register scheduling, reduced the multiplication cycles of \HQC{1} and \HQC{3} by $33.4\%$ and $27.2\%$, respectively, over the preceding work~\cite{MyeonghoonIOT}. In parallel, Chen et al.~\cite{TCHES:CCC26} reinterpreted the Encode step of FAFFT as a ring homomorphism and proposed an FAFFT+CRT method that multiplies without padding, using only a single $32{,}768$-point FAFFT and an auxiliary multiplication over a small residue ring. This work builds on the implementation of Jang et al.~\cite{jlk25}, integrates the optimizations of this concurrent work~\cite{TCHES:CCC26}, and adopts both implementations as performance baselines.


\section{Preliminaries}
\label{sec:preliminaries}
\subsection{Notation}

Let $\mathbb{F}_2$ be the binary field. A length-$n$ vector over $\mathbb{F}_2$ is identified interchangeably with a polynomial of the ring $\mathcal{R} = \mathbb{F}_2[X]/(X^n - 1)$, and the product of two such elements is defined as their product in $\mathcal{R}$. Vectors and polynomials are written in lowercase boldface ($\mathbf{u}, \mathbf{v}$). For a set $S$, $s \stackrel{\$}{\gets} S$ denotes that $s$ is drawn at random from $S$, and $s \stackrel{\$}{\gets} S$ from $\mathit{seed}$ denotes pseudorandom sampling determined by $\mathit{seed}$. Let $\omega(\cdot)$ denote the Hamming weight of a vector, and for a positive integer $\omega$ let $\mathcal{R}_\omega := \{\mathbf{v} \in \mathcal{R} \mid \omega(\mathbf{v}) = \omega\}$ be the set of vectors of weight $\omega$. For $\mathbf{v} = (v_0, \dots, v_{n-1})$ with index $0$ as the least significant bit and for $0 \le n' \le n$, define $\mathsf{Trun}(\mathbf{v}, n')$ as the function that discards the top $n - n'$ bits and keeps only the low $n'$ bits.

\subsection{Hamming Quasi-Cyclic (HQC)}

HQC is an IND-CCA2 KEM based on the hardness of syndrome decoding for quasi-cyclic codes, obtained by applying the Fujisaki--Okamoto transform with implicit rejection to an IND-CPA-secure public-key encryption scheme (HQC-PKE)~\cite{hqc-spec}. Its main components are given in Algorithms~\ref{alg:HQC-PKE} and~\ref{alg:HQC-KEM}, where $\mathtt{H}, \mathtt{G}, \mathtt{J}$ are hash functions. Error correction uses a concatenated code $\mathcal{C}$ of length $n_1 n_2$ that serially combines a Reed-Muller (RM) code of length $n_2$ and a Reed-Solomon (RS) code of length $n_1$.

\begin{algorithm}[h!]
\caption{Main components in \textsf{HQC-PKE}~\cite{hqc-spec}.}
\label{alg:HQC-PKE}
\linespread{1.2}\footnotesize
\begin{algorithmic}[1]
\State \textbf{Global parameters:} $(n, n_1, n_2, \omega, \omega_r, \omega_e, \mathcal{C})$
\algstore{hqcpke}
\end{algorithmic}
\begin{minipage}[t]{0.53\textwidth}
\begin{algorithmic}[1]
\algrestore{hqcpke}
\State \textsf{HQC-PKE.Keygen}$(\mathit{seed}_{\mathsf{PKE}})$
\State \quad $(\mathit{seed}_{\mathsf{dk}}, \mathit{seed}_{\mathsf{ek}}) \gets I(\mathit{seed}_{\mathsf{PKE}})$
\State \quad $(\mathbf{y}, \mathbf{x}) \stackrel{\$}{\gets} \mathcal{R}_\omega \times \mathcal{R}_\omega$ from $\mathit{seed}_{\mathsf{dk}}$
\State \quad $\mathbf{h} \stackrel{\$}{\gets} \mathcal{R}$ from $\mathit{seed}_{\mathsf{ek}}$
\State \quad $\mathbf{s} \gets \mathbf{x} + \mathbf{h} \cdot \mathbf{y}$
\State \quad \textbf{return} $\mathsf{ek}_{\mathsf{PKE}} \gets (\mathit{seed}_{\mathsf{ek}}, \mathbf{s})$, $\mathsf{dk}_{\mathsf{PKE}} \gets \mathit{seed}_{\mathsf{dk}}$
\algstore{hqcpke}
\end{algorithmic}
\end{minipage}\hfill
\begin{minipage}[t]{0.45\textwidth}
\begin{algorithmic}[1]
\algrestore{hqcpke}
\State \textsf{HQC-PKE.Encrypt}$(\mathsf{ek}_{\mathsf{PKE}}, \mathbf{m}, \theta)$
\State \quad $\mathbf{h} \stackrel{\$}{\gets} \mathcal{R}$ from $\mathit{seed}_{\mathsf{ek}}$
\State \quad $(\mathbf{r}_2, \mathbf{e}, \mathbf{r}_1) \stackrel{\$}{\gets} (\mathcal{R}_{\omega_r}, \mathcal{R}_{\omega_e}, \mathcal{R}_{\omega_r})$ from $\theta$
\State \quad $\mathbf{u} \gets \mathbf{r}_1 + \mathbf{h} \cdot \mathbf{r}_2$
\State \quad $\mathbf{v} \gets \mathcal{C}.\mathsf{Encode}(\mathbf{m}) + \mathsf{Trun}(\mathbf{s} \cdot \mathbf{r}_2 + \mathbf{e}, n_1 n_2)$
\State \quad \textbf{return} $\mathbf{c}_{\mathsf{PKE}} \gets (\mathbf{u}, \mathbf{v})$
\algstore{hqcpke}
\end{algorithmic}
\end{minipage}
\par\vspace{3pt}\hrule\vspace{3pt}
\begin{algorithmic}[1]
\algrestore{hqcpke}
\State \textsf{HQC-PKE.Decrypt}$(\mathsf{dk}_{\mathsf{PKE}}, \mathbf{c}_{\mathsf{PKE}} = (\mathbf{u}, \mathbf{v}))$
\State \quad $\mathbf{y} \stackrel{\$}{\gets} \mathcal{R}_\omega$ from $\mathit{seed}_{\mathsf{dk}}$
\State \quad \textbf{return} $\mathbf{m} \gets \mathcal{C}.\mathsf{Decode}(\mathbf{v} - \mathsf{Trun}(\mathbf{u} \cdot \mathbf{y}, n_1 n_2))$
\end{algorithmic}
\end{algorithm}

\begin{algorithm}[h!]
\caption{Main components in \textsf{HQC-KEM}~\cite{hqc-spec}.}
\label{alg:HQC-KEM}
\linespread{1.2}\footnotesize
\begin{minipage}[t]{0.5\textwidth}
\begin{algorithmic}[1]
\State \textsf{HQC-KEM.Keygen}$()$
\State \quad $\mathit{seed}_{\mathsf{KEM}} \stackrel{\$}{\gets} \{0,1\}^{|\mathit{seed}|}$
\State \quad $(\mathit{seed}_{\mathsf{PKE}}, \sigma) \gets \mathtt{XOF}(\mathit{seed}_{\mathsf{KEM}})$
\State \quad $(\mathsf{ek}_{\mathsf{PKE}}, \mathsf{dk}_{\mathsf{PKE}}) \gets \textsf{HQC-PKE.Keygen}(\mathit{seed}_{\mathsf{PKE}})$
\State \quad $\mathsf{ek}_{\mathsf{KEM}} \gets \mathsf{ek}_{\mathsf{PKE}}$
\State \quad $\mathsf{dk}_{\mathsf{KEM}} \gets (\mathsf{ek}_{\mathsf{KEM}}, \mathsf{dk}_{\mathsf{PKE}}, \sigma, \mathit{seed}_{\mathsf{KEM}})$
\State \quad \textbf{return} $\mathsf{ek}_{\mathsf{KEM}}$, $\mathsf{dk}_{\mathsf{KEM}}$
\algstore{hqckem}
\end{algorithmic}
\end{minipage}\hfill
\begin{minipage}[t]{0.5\textwidth}
\begin{algorithmic}[1]
\algrestore{hqckem}
\State \textsf{HQC-KEM.Encaps}$(\mathsf{ek}_{\mathsf{KEM}})$
\State \quad $\mathbf{m} \stackrel{\$}{\gets} \{0,1\}^{k}$, \; $\mathit{salt} \stackrel{\$}{\gets} \{0,1\}^{|\mathit{salt}|}$
\State \quad $(K, \theta) \gets \mathtt{G}(\mathtt{H}(\mathsf{ek}_{\mathsf{KEM}}) \,\|\, \mathbf{m} \,\|\, \mathit{salt})$
\State \quad $\mathbf{c}_{\mathsf{PKE}} \gets \textsf{HQC-PKE.Encrypt}(\mathsf{ek}_{\mathsf{KEM}}, \mathbf{m}, \theta)$
\State \quad \textbf{return} $K$, $\mathbf{c}_{\mathsf{KEM}} \gets (\mathbf{c}_{\mathsf{PKE}}, \mathit{salt})$
\algstore{hqckem}
\end{algorithmic}
\end{minipage}
\par\vspace{3pt}\hrule\vspace{3pt}
\begin{algorithmic}[1]
\algrestore{hqckem}
\State \textsf{HQC-KEM.Decaps}$(\mathsf{dk}_{\mathsf{KEM}}, \mathbf{c}_{\mathsf{KEM}})$
\State \quad Parse $\mathsf{dk}_{\mathsf{KEM}} = (\mathsf{ek}_{\mathsf{KEM}}, \mathsf{dk}_{\mathsf{PKE}}, \sigma, \mathit{seed}_{\mathsf{KEM}})$, \; $\mathbf{c}_{\mathsf{KEM}} = (\mathbf{c}_{\mathsf{PKE}}, \mathit{salt})$
\State \quad $\mathbf{m}' \gets \textsf{HQC-PKE.Decrypt}(\mathsf{dk}_{\mathsf{PKE}}, \mathbf{c}_{\mathsf{PKE}})$
\State \quad $(K', \theta') \gets \mathtt{G}(\mathtt{H}(\mathsf{ek}_{\mathsf{KEM}}) \,\|\, \mathbf{m}' \,\|\, \mathit{salt})$
\State \quad $\mathbf{c}'_{\mathsf{PKE}} \gets \textsf{HQC-PKE.Encrypt}(\mathsf{ek}_{\mathsf{KEM}}, \mathbf{m}', \theta')$
\State \quad $\bar{K} \gets \mathtt{J}(\mathtt{H}(\mathsf{ek}_{\mathsf{KEM}}) \,\|\, \sigma \,\|\, \mathbf{c}_{\mathsf{KEM}})$
\State \quad \textbf{if} $\mathbf{m}' = \bot$ \textbf{ or } $\mathbf{c}'_{\mathsf{PKE}} \neq \mathbf{c}_{\mathsf{PKE}}$ \textbf{ then } $K' \gets \bar{K}$
\State \quad \textbf{return} $K'$
\end{algorithmic}
\end{algorithm}

HQC involves (i) binary polynomial multiplication in the ring $\mathcal{R}$ ($\mathbf{h}\cdot\mathbf{r}_2$, $\mathbf{s}\cdot\mathbf{r}_2$, and $\mathbf{u}\cdot\mathbf{y}$) and (ii) the sampling that generates the fixed-weight vectors ($\mathbf{x}, \mathbf{y}, \mathbf{r}_1, \mathbf{r}_2, \mathbf{e}$) and the public vector $\mathbf{h}$ from a SHAKE-based XOF. Together with hashing, these two operations dominate the performance of HQC, and this paper targets the multiplication and the sampling for optimization.

\subsection{Frobenius Additive FFT (FAFFT)}

The product of two elements $\mathbf{a}, \mathbf{b}$ of the ring $\mathcal{R} = \mathbb{F}_2[X]/(X^n - 1)$ can be computed with an additive FFT. We first take $l$ to be the smallest power of two greater than $2n-2$, and regard the coefficients as elements of a tower field $\mathbb{F}_{2^m}$ with $2^m \ge l$ and $m$ a power of two, viewing $\mathbf{a}, \mathbf{b}$ as polynomials over $\mathbb{F}_{2^m}$. We then (i) evaluate the two polynomials at $l$ points with the additive FFT, (ii) multiply them pointwise, (iii) recover the product polynomial with the inverse additive FFT, and (iv) reduce it into $\mathcal{R}$. Since all input coefficients lie in $\mathbb{F}_2$, the recovered product also has its coefficients in $\mathbb{F}_2$, so mapping it back from $\mathbb{F}_{2^m}$ to $\mathbb{F}_2$ requires no additional computation~\cite{MyeonghoonIOT}.

The additive FFT~\cite{gao2010additive, cantor1989arithmetical, lin2016fft} recursively evaluates a polynomial $f \in \mathbb{F}_{2^m}[x]$ at many points. For a Cantor basis $\{v_0, \dots, v_{m-1}\}$ of $\mathbb{F}_{2^m}$ over $\mathbb{F}_2$, let $W_k = \langle v_0, \dots, v_{k-1} \rangle$ ($0 \le k \le m$, $|W_k| = 2^k$, $W_0 = \{0\}$) be the associated subspaces, and define the subspace polynomials \(s_0(x) = x, s_{i+1}(x) = s_i(x)^2 + s_i(x).\) Each $s_i$ is linear over $\mathbb{F}_2$ with $\deg s_i = |W_i| = 2^i$, and $s_k(x) = \prod_{a \in W_k}(x - a)$ vanishes on $W_k$. The additive FFT then consists of a butterfly recursion that splits $f$ into two halves using $s_i$, whose key operation at each butterfly step is
\[
  h_0(X) \gets p_0(X) + s_i(\alpha)\cdot p_1(X), \qquad h_1(X) \gets h_0(X) + p_1(X),
\]
where $s_i(\alpha)$ is a fixed constant determined at each butterfly node. Since all operations are carried out over $\mathbb{F}_{2^m}$, they reduce to XORs and carry-less multiplications. In particular, with a Cantor basis, $s_k(x) = x^{2^k} + x$ becomes sparse when $k$ is a power of two, so the low-level butterflies reduce to simple bit operations.

The Frobenius additive FFT (FAFFT) reduces the number of evaluation points using the Frobenius map $\phi_2 : a \mapsto a^2$~\cite{li2018frobenius}. When the input polynomial $f$ has coefficients in $\mathbb{F}_2$,\(f(\phi_2(a)) = f(a^2) = f(a)^2 = \phi_2\!\left(f(a)\right)\) holds, so once the value $f(a)$ at one point is given, the values $f(a^2), f(a^4), \dots$ are also fixed. For a set $\Sigma$, let $\mathrm{Ord}_{\phi_2}(\Sigma)$ be the smallest integer $j$ with $\phi_2^{\,j}(\Sigma) = \Sigma$, and call $\Sigma$ a Frobenius partition of the evaluation domain $\Omega$ when
\[
  \Omega = \Sigma \,\cup\, \phi_2(\Sigma) \,\cup\, \cdots \,\cup\, \phi_2^{\,j-1}(\Sigma)
\]
is a disjoint union. Li et al.~\cite{li2018frobenius} showed that, for the Cantor basis, $\phi_2(W_i) = W_i$, and for $k = \log_2 l - \log_2 m$ the set $\Sigma = v_{k + m/2} + W_k$ is a Frobenius partition with $\mathrm{Ord}_{\phi_2}(\Sigma) = m$ and $|\Sigma| = 2^k = l/m$. Evaluating $f$ at the $l$ points of $\Omega$ therefore reduces to evaluating at the $l/m$ representative points of $\Sigma$, and the values at the remaining points are already fixed by the Frobenius relation and are not computed separately. Consequently, the product $\mathbf{c}$ of two polynomials $\mathbf{a}, \mathbf{b}$ is obtained by multiplying $\mathsf{FAFFT}(\mathbf{a})$ and $\mathsf{FAFFT}(\mathbf{b})$ pointwise at the $l/m$ evaluation points and then applying the inverse transform $\mathsf{FAFFT}^{-1}$.

In the Cortex-M4 implementation, this computation is carried out in a bit-sliced form over the tower field $\mathbb{F}_2 \subset \mathbb{F}_{2^2} \subset \cdots \subset \mathbb{F}_{2^{32}}$, and the additions in the butterfly are handled as XORs between bit planes. Among the field multiplications in the butterfly, multiplication by the constant fixed at each level is a fixed $\mathbb{F}_2$-linear map, so it is implemented as a dedicated straight-line XOR sequence~\cite{MyeonghoonIOT}, whereas multiplication by the constant that varies with the evaluation point is performed as a general tower-field multiplication. The former is straight-line code consisting of XORs and register moves, and this paper optimizes its instruction scheduling in \Sref{sec:vmov}.

However, the polynomial length $n$ of HQC is slightly larger than a power of two, so applying the FAFFT directly incurs the overhead of zero-padding up to the next power of two, which is $65{,}536$ for \HQC{1}. To reduce this, a method was proposed that uses the Chinese Remainder Theorem (CRT) to decompose the polynomial ring into a product of a ring suited to the FAFFT and a small residue ring, so that the multiplication is handled with a single FAFFT of half that size and an auxiliary multiplication in the small residue ring~\cite{jlk25, TCHES:CCC26}. Among these, this paper uses the implementation of Jang et al.~\cite{jlk25}, with the techniques of Chen et al.~\cite{TCHES:CCC26} integrated, as the basis for its own optimizations.
\subsection{ARM Cortex-M4}

The ARM Cortex-M4 is a 32-bit embedded processor based on the ARMv7E-M architecture. It provides the Thumb-2 instruction set and DSP extensions such as single-cycle multiplication and multiply-accumulate (MAC), but it has no carry-less multiplication instruction such as x86's \texttt{PCLMULQDQ} or RISC-V's \texttt{clmul}. The Thumb-2 instruction set supports predicated execution through \texttt{IT} (If-Then) blocks, which lets a condition flag control whether an instruction takes effect without branching. When the condition is false, the instruction still occupies its execution slot but writes no result, which is useful for constant-time implementations that avoid secret-dependent branches. 

The ARM Cortex-M4 has 16 GP registers, \texttt{r0}--\texttt{r15}. Of these, \texttt{r13} (SP) and \texttt{r15} (PC) are reserved, leaving 14 available for general use. The optional floating-point unit (FPv4-SP) additionally provides 32 single-precision registers \texttt{s0}--\texttt{s31}, which can serve not only for floating-point operations but also as extra storage. However, data movement between the GP registers and the floating-point registers is possible only through the \texttt{vmov} instruction and incurs a cycle cost per move. This \texttt{vmov} cost is the key overhead of the bit-sliced FAFFT butterfly, which this paper minimizes in \Sref{sec:vmov}.

\section{Polynomial Multiplication Optimization}
\label{sec:vmov}

This section presents our optimizations to the polynomial multiplication that dominates the cost of HQC. Building on the FAFFT-CRT implementation of Jang et al.~\cite{jlk25}, we incorporate the optimizations of Chen et al.~\cite{TCHES:CCC26} and lower the FAFFT modulus weight further for \HQC{1} (\Sref{sec:polymul-combine}). We then minimize the VMOV overhead of the bit-sliced butterfly through register allocation and XOR reordering (\Sref{sec:vmov-min}), and report the resulting multiplication cycles in \Sref{sec:polymul-eval}.

\subsection{Combining FAFFT-CRT Multiplication Techniques}
\label{sec:polymul-combine}

The dominant operation of HQC is the multiplication $\mathbf{c} = \mathbf{a}\cdot\mathbf{b}$ in $\mathcal{R} = \mathbb{F}_2[x]/(x^n-1)$, where $n = 2^d + r$ for a small $r$. Two concurrent works, Jang et al.~\cite{jlk25} and Chen et al.~\cite{TCHES:CCC26}, both compute it by combining the FAFFT with the CRT, splitting $\mathcal{R}$ into a $2^d$-point FAFFT component and a small residue ring handled by a radix-16 multiplication, which avoids the zero-padding to $2^{d+1}$ that a direct transform would otherwise require.

We take the implementation of Jang et al.~\cite{jlk25} as our base and incorporate three optimizations of Chen et al.~\cite{TCHES:CCC26}. First, a \emph{truncated basis conversion}: as each operand fills only part of the transform, the conversion produces the occupied prefix instead of the full $2^d$ coefficients, namely $16{,}384 + 2{,}048$ for the $32{,}768$-point transform of \HQC{1} and $32{,}768 + 4{,}096$ for the $65{,}536$-point transform of \HQC{3}. Second, a \emph{low-weight FAFFT modulus} that keeps the sparse-dense reconstruction cheap; here we go beyond Chen et al. and find a sparser modulus for \HQC{1} (\Sref{sec:polymul-basis}). Third, \emph{sharing} the forward transform of the ephemeral operand $\mathbf{r}_2$, which is common to the products $\mathbf{h}\cdot\mathbf{r}_2$ and $\mathbf{s}\cdot\mathbf{r}_2$ in encapsulation and decapsulation, so that it is computed only once.

The residue ring itself is handled by a radix-16 polynomial multiplication, and this is where the two prior works differ. For \HQC{1} and \HQC{3}, the residue product occupies $2{,}570$ and $6{,}166$ bits, that is $\lceil 2570/32 \rceil = 81$ and $\lceil 6166/32 \rceil = 193$ 32-bit words. Jang et al.~\cite{jlk25} multiply the residue at these exact sizes of $81$ and $193$, whereas Chen et al.~\cite{TCHES:CCC26} zero-pad them to the implementation-friendly sizes of $96$ and $256$ while skipping the high-order words that the truncated product does not need. For \HQC{1} the padding from $81$ to $96$ is small, so the truncated multiplication of Chen et al. is faster, but for \HQC{3} the padding from $193$ to $256$ is large enough that the exact multiplication of Jang et al. is faster. We therefore take the residue multiplication of Chen et al. for \HQC{1} and that of Jang et al. for \HQC{3}.

These optimizations apply to \HQC{1} and \HQC{3}, whose parameters admit the CRT split. \HQC{5} uses a pure double-size FAFFT with no residue ring.

Beyond the multiplication, we also adopt the RS/RM decoder of Chen et al.~\cite{TCHES:CCC26} for the decapsulation decoding. Decoding is not a multiplication, but it is another component where the base implementation of Jang et al.~\cite{jlk25} and that of Chen et al. differ. The Reed-Solomon step of HQC recovers the error-locator polynomial from the syndromes, and Chen et al. replace the Berlekamp-Massey solver of the base decoder with an Extended Euclidean Algorithm (EEA). The regular data flow of the EEA aligns with their $\mathbb{F}_{256}$-optimized arithmetic, implemented as unrolled radix-16 matrix-vector products, which makes the decoder faster on the Cortex-M4 at the cost of the larger code discussed in \Sref{sec:evaluation}. Since decoding occurs only in decapsulation, adopting it further reduces the decapsulation cost.

\subsubsection{A Low-Weight FAFFT Modulus}
\label{sec:polymul-basis}

In the Hybrid FAFFT-CRT of \Sref{sec:polymul-combine}, the product is reconstructed as $C(x) = C_1(x) + t(x)\cdot\mathcal{P}(x)$, where $\mathcal{P}(x)$ is the vanishing polynomial of the FAFFT evaluation domain. Because $\mathcal{P}(x)$ is fixed, public, and sparse, the reconstruction term $t(x)\cdot\mathcal{P}(x)$ is a constant-time sparse-dense multiplication whose cost is proportional to the Hamming weight $\omega(\mathcal{P})$; the sparser $\mathcal{P}(x)$ is, the cheaper the reconstruction.

The evaluation set that determines $\mathcal{P}(x)$ is not unique, so $\omega(\mathcal{P})$ depends on this choice. Earlier additive-FFT work did not optimize it, leaving $\omega(\mathcal{P}) = 501$ for \HQC{1}~\cite{jlk25}, whereas Chen et al.~\cite{TCHES:CCC26} minimized it to $165$ for \HQC{1} and $199$ for \HQC{3}. We search for the minimizing evaluation set independently for each parameter: for \HQC{3} the value $199$ is already minimal and we retain it, while for \HQC{1} we find an evaluation set with $\omega(\mathcal{P}) = 109$, below the previous $165$ (a $34\%$ reduction), which directly lowers the reconstruction cost. \HQC{5} uses a pure double-size FAFFT with no residue ring, so this optimization does not apply. The new \HQC{1} evaluation set also changes the butterfly's fixed-constant multiplications and hence their XOR sequences; since these have no prior implementation, the \HQC{1} reference of \Sref{sec:vmov-min} is reconstructed accordingly (marked with an asterisk).

\subsection{VMOV Minimization via Register Allocation and XOR Reordering}
\label{sec:vmov-min}

We show the limitation of the existing framework that minimizes only the XOR count in the assembly implementation of FAFFT butterfly multiplication operations, and propose a novel framework that optimizes the number of data movements between GP registers and VFP registers, namely VMOV instructions, through register allocation and XOR reordering.

\subsubsection{Motivation}
\label{sec:vmov-problem}

In the butterfly stage of FAFFT, terms of the form $s_i(\alpha)\cdot p_1(X)$ are computed repeatedly. When $m=32$, $s_i(\alpha)$ decomposes as $s_i(\alpha)=w+v$, where $w \in W_k$ and $v \in \{v_{17},\dots,v_{k+16}\}$ is fixed per stage and hence common to all nodes of that stage. Therefore, it can be written as
\[
s_i(\alpha)\cdot p_1(X)=w\cdot p_1(X)+v\cdot p_1(X).
\]
Since $k<16$, we have $W_k \subseteq W_{16}=\mathbb{F}_{2^{16}}$, so $w$ lies in a proper subfield of $\mathbb{F}_{2^{32}}$ and $w\cdot p_1(X)$ can be computed by multiplications in that subfield rather than in $\mathbb{F}_{2^{32}}$~\cite{chen2021optimizing}. In $v\cdot p_1(X)$, on the other hand, $v$ is a fixed constant, so this operation is a multiplication by a fixed constant in $\mathbb{F}_{2^{32}}$. Lee et al.~\cite{MyeonghoonIOT} observed that this is a $32 \times 32$ $\mathbb{F}_2$-linear transformation mapping a 32-bit input vector to a 32-bit output vector, and can therefore be represented as a matrix-vector multiplication implementable using only XOR operations.

The butterfly operations discussed in this section are straight-line implementations of multiplications by the fixed constants $v_{17}, \dots, v_{28}$. Although the corresponding linear transformations are expressed using EOR instructions, the implementation must handle 32 state variables with only 14 available GP registers. Values are therefore transferred between GP registers and dedicated VFP registers using VMOV instructions, and we use the sum of the EOR and VMOV counts as the implementation cost. This register pressure motivates the following observations.

\paragraph{Observation 1: Half of the instructions are data movement rather than arithmetic.}
As shown in \Tref{tab:vmov-merged}, across the 12 butterfly assemblies of Jang et al.~\cite{jlk25}, the total number of EOR instructions is $2{,}034$, while the total number of VMOV instructions is $1{,}992$. In other words, the number of XOR instructions implementing the $\mathbb{F}_2$-linear transformations is almost the same as the number of data movement instructions. For the implementation corresponding to each fixed constant $v$, VMOV instructions account for 47--51\% of the total instructions. In particular, for three constants ($v_{23}, v_{26}, v_{27}$), the number of VMOV instructions exceeds the number of EOR instructions. For example, the implementation for $v_{26}$ consists of $179$ EOR instructions and $187$ VMOV instructions, so it spends more clock cycles on data movement than on the fixed-constant multiplication itself. These results show that, in the actual Cortex-M4 implementation, roughly half of the clock cycles correspond not to the $\mathbb{F}_2$-linear transformation itself but to the overhead caused by register pressure.

\paragraph{Observation 2: Reducing the XOR count does not always reduce the instruction cost.}
As in Lee et al.~\cite{MyeonghoonIOT} and Jang et al.~\cite{jlk25}, we also apply the framework of Xiang et al.~\cite{xiang2020linear} independently to each of the 12 fixed-constant multiplications to further search for shorter XOR sequences. As a result, we find new sequences with fewer XOR operations for five constants ($v_{18}, v_{19}, v_{20}, v_{24}, v_{28}$). We convert these five sequences into assembly implementations with the same functional structure as the implementation of Jang et al.~\cite{jlk25} and compare them with the original implementation. Our search results are shown in \Tref{tab:xorres-motiv}. The number of EOR instructions decreases by 1--3 for each constant, but the number of VMOV instructions increases by 14--36. As a result, all five implementations have larger total instruction counts than those of Jang et al.

\begin{table}[!h]
\centering
\small
\caption{Results of converting shorter XOR sequences found using the framework of Xiang et al.~\cite{xiang2020linear} into assembly.}
\label{tab:xorres-motiv}
\setlength{\tabcolsep}{4.5pt}
\begin{tabular}{lccccccc}
\toprule
 & \multicolumn{3}{c}{Jang et al.~\cite{jlk25}}
 & \multicolumn{3}{c}{New XOR sequence} & \\
\cmidrule(lr){2-4} \cmidrule(lr){5-7}
$v_i$ & \#EOR & \#VMOV & Total & \#EOR & \#VMOV & Total & $\Delta$Total \\
\midrule
$v_{18}$ & 153 & 139 & 292 & 150 & 157 & 307 & $+15$ \\
$v_{19}$ & 166 & 161 & 327 & 165 & 175 & 340 & $+13$ \\
$v_{20}$ & 172 & 167 & 339 & 170 & 186 & 356 & $+17$ \\
$v_{24}$ & 180 & 171 & 351 & 178 & 207 & 385 & $+34$ \\
$v_{28}$ & 183 & 173 & 356 & 181 & 196 & 377 & $+21$ \\
\bottomrule
\end{tabular}
\end{table}

These results show that reducing the XOR count does not necessarily reduce the total instruction count. Therefore, optimization on the Cortex-M4 must account for the VMOV cost caused by register pressure. Motivated by these observations, we minimize VMOV instructions by reconsidering the execution order and register allocation while preserving the resulting matrix $M$.

\subsubsection{XOR Sequences and the Implementation Cost Model}
\label{sec:vmov-cost-model}

Each butterfly function is a linear transformation $M \in \mathrm{GL}(m, \mathbb{F}_2)$ with $m=32$, which decomposes into a product of type-1 and type-3 elementary matrices. A type-3 matrix $E(i + j)$ adds row $j$ to row $i$ of the identity and corresponds to an in-place XOR, $x_i \leftarrow x_i \oplus x_j$. A type-1 matrix $E(i \leftrightarrow j)$ swaps rows $i$ and $j$, exchanging the positions of two variables. Thus $M$ can be written as
\begin{equation}
   M \;=\; E(i_n + j_n) \cdots E(i_1 + j_1)\,
           E(i'_s \leftrightarrow j'_s) \cdots E(i'_1 \leftrightarrow j'_1)
   \;\in\; \mathrm{GL}(m, \mathbb{F}_2).
   \label{eq:M-of-L}
\end{equation}

A type-1 elementary matrix only relabels the register holding each variable and therefore incurs no cost in the actual implementation. The type-3 part of~\eqref{eq:M-of-L} corresponds to an ordered sequence of in-place XOR operations applied to the state variables $x_0,\dots,x_{m-1}$, from $x_{i_1}\leftarrow x_{i_1}\oplus x_{j_1}$ through $x_{i_n}\leftarrow x_{i_n}\oplus x_{j_n}$ in turn. We call such a sequence of operations an \emph{XOR sequence} and denote it by $L$. Its length $n$ corresponds directly to the number of EOR instructions needed to implement the transformation. 
The existing framework of Xiang et al.~\cite{xiang2020linear} takes exactly this count as its objective and, among the decompositions that preserve $M$, searches for an XOR sequence with fewer type-3 elementary matrices under a set of reduction rules.

In the actual implementation, an EOR instruction can operate only on operands held in GP registers, so executing $x_i\leftarrow x_i\oplus x_j$ requires both the source $x_j$ and the destination $x_i$ to reside there. An XOR sequence, however, specifies only the source and destination of each operation, not the registers in which they are placed, and the register constraint of \Sref{sec:vmov-problem} prevents all state variables from residing in GP registers simultaneously; a register-allocation policy is therefore needed to decide which variables to keep at each point. If a required variable is absent, its VFP copy must be brought in, incurring one load VMOV, and if all GP registers are occupied, a resident variable must first be evicted. We call a resident variable \emph{dirty} if its value has been updated since its last load and \emph{clean} otherwise, writing $\mathcal{D}$ for the set of dirty variables. Evicting a clean variable is free, since its VFP copy is still valid, whereas evicting a dirty variable costs one spill VMOV to write its latest value back; the VMOV count of an instruction sequence is therefore the number of loads plus the number of dirty-variable spills.

The EOR count $n$ is fixed once the XOR sequence is chosen, whereas the VMOV count varies with the register-allocation policy. The only variable term of the implementation cost---the sum of the EOR and VMOV counts---is thus the VMOV count, and the same XOR sequence may translate into instruction sequences of different lengths under different policies. Reducing the actual instruction count therefore requires a register-allocation policy that accounts for the cost of moving values between GP and VFP registers.

\subsubsection{Reducing VMOV Costs via Register Allocation}
\label{sec:vmov-theory}

We now describe how to allocate registers for a fixed XOR sequence so as to minimize the VMOV count. We first recall the farthest-first policy used in prior work and expose its limitation, and then propose our Spill-Penalized Farthest-First policy.

\paragraph{The Farthest-First Policy and Its Limitation}
\label{sec:vmov-theory-count}

When the XOR sequence is fixed, the order of operations is known in advance, which enables an eviction policy that reduces unnecessary reloads. The prior works of Lee et al.~\cite{MyeonghoonIOT} and Jang et al.~\cite{jlk25} use the farthest-first (FF) policy, which, whenever all GP registers are occupied, evicts the variable whose next use lies farthest in the future. By keeping variables that will soon be reused in GP registers, this policy aims to reduce subsequent load VMOVs.

Let $\operatorname{nextuse}(x,t)$ denote the first time after step $t$ at which variable $x$ is used again, set to $\infty$ if it is never used again. Let $\mathcal{R}_t$ denote the set of eviction candidates at step $t$, excluding the source and destination of the XOR operation about to be executed. The FF policy then selects the eviction target as
\[
x_u=
\operatorname*{arg\,max}_{x\in \mathcal{R}_t}
\operatorname{nextuse}(x,t).
\]
That is, FF evicts the candidate whose next use is farthest away. Since the entire XOR sequence is known in advance, each variable's next-use time can be computed exactly.

However, FF decides the eviction target solely from the next-use time and thus ignores the cost difference due to a variable's dirty status. Evicting a clean variable incurs no VMOV, whereas evicting a dirty variable incurs one spill VMOV to write its latest value back to a VFP register. Consequently, when candidates share the same next-use time, or when a dirty variable's next use is only slightly farther than a clean one's, evicting the dirty variable can make the spill cost outweigh the gain from reduced reloads.

\Tref{tab:ff-limit} illustrates this limitation with $K=3$ available GP registers. After the second operation, $x_0$, $x_1$, and $x_2$ are resident, with $x_0$ dirty from having been the destination of both operations and $x_1$, $x_2$ clean. The third operation $x_1\leftarrow x_1\oplus x_3$ then requires loading $x_3$, so one resident variable must be evicted; excluding the destination $x_1$ leaves $x_0$ and $x_2$, both of which are reused at the fourth operation and hence share the same next-use time. FF breaks this tie without regard to dirty status and may thus evict $x_0$, spilling its latest value and reloading it one operation later, which brings the total to $6$ VMOVs. Even though FF accounts for the order of future uses, it does not reflect the eviction cost that depends on dirty status.

\begin{table}[!h]
\centering
\caption{Register allocation with plain FF under an unfavorable tie-breaking.}
\label{tab:ff-limit}
\small
\resizebox{\linewidth}{!}{%
\begin{tabular}{c|l|cc|cc|l|c}
\toprule
 & & \multicolumn{2}{c|}{Before} & \multicolumn{2}{c|}{After} & & \\
$t$ & XOR operation & Dirty & Clean & Dirty & Clean & Required \texttt{VMOV} & \#\texttt{VMOV} \\
\midrule
1 & $x_0\leftarrow x_0\oplus x_1$ & $\emptyset$ & $\emptyset$ & $\{x_0\}$ & $\{x_1\}$ & load $x_0$, load $x_1$ & 2 \\
2 & $x_0\leftarrow x_0\oplus x_2$ & $\{x_0\}$ & $\{x_1\}$ & $\{x_0\}$ & $\{x_1,x_2\}$ & load $x_2$ & 1 \\
3 & $x_1\leftarrow x_1\oplus x_3$ & $\{x_0\}$ & $\{x_1,x_2\}$ & $\{x_1\}$ & $\{x_2,x_3\}$ & spill $x_0$, load $x_3$ & 2 \\
4 & $x_0\leftarrow x_0\oplus x_2$ & $\{x_1\}$ & $\{x_2,x_3\}$ & $\{x_0,x_1\}$ & $\{x_2\}$ & load $x_0$ (drop $x_3$) & 1 \\
5 & $x_1\leftarrow x_1\oplus x_2$ & $\{x_0,x_1\}$ & $\{x_2\}$ & $\{x_0,x_1\}$ & $\{x_2\}$ & - & 0 \\
\bottomrule
\end{tabular}
}
\end{table}

\paragraph{The Spill-Penalized Dirty-Aware Farthest-First Policy}
\label{sec:SPF}

To reflect in victim selection the spill cost that FF overlooks, we propose the Spill-Penalized Dirty-Aware Farthest-First (SPF) policy. SPF retains FF's $\operatorname{nextuse}$-based selection criterion but incorporates into the score the spill cost that may arise when a dirty variable is evicted. For each eviction candidate $x\in\mathcal{R}_t$ at step $t$, we define the following score.
\[
S_p(x,t)
=
\operatorname{nextuse}(x,t) - p\cdot \mathbf{1}[x\in\mathcal{D}],
\]
Here $\mathbf{1}[x\in\mathcal{D}]$ is $1$ if $x$ is dirty and $0$ otherwise, and $p$ is the penalty imposed on evicting a dirty variable. SPF evicts the candidate with the largest $S_p$, breaking ties in favor of clean variables. Thus $p=0$ is not plain FF but FF augmented with a dirty-aware tie-break. When $p>0$, a dirty variable's score is reduced, so the clean variable is evicted even if the dirty one is used later, as long as the difference is at most $p$; that is, $p$ is the loss in next-use distance that SPF accepts in order to avoid a dirty spill.

\Tref{tab:SPF-example} applies SPF to the same XOR sequence as the FF example. Just before the third operation, the candidates $x_0$ and $x_2$ have equal next-use times, but where FF breaks the tie by evicting the dirty $x_0$ and incurring a spill (\Tref{tab:ff-limit}), SPF evicts the clean $x_2$. The reload count is unchanged---$x_2$ is reloaded at the fourth operation, just as FF reloads $x_0$---so avoiding the spill reduces the total VMOV count from $6$ to $5$.

\begin{table}[!h]
\centering
\caption{Register allocation with SPF.}
\label{tab:SPF-example}
\small
\resizebox{\linewidth}{!}{%
\begin{tabular}{c|l|cc|cc|l|c}
\toprule
 & & \multicolumn{2}{c|}{Before} & \multicolumn{2}{c|}{After} & & \\
$t$ & XOR operation & Dirty & Clean & Dirty & Clean & Required \texttt{VMOV} & \#\texttt{VMOV} \\
\midrule
1 & $x_0\leftarrow x_0\oplus x_1$ & $\emptyset$ & $\emptyset$ & $\{x_0\}$ & $\{x_1\}$ & load $x_0$, load $x_1$ & 2 \\
2 & $x_0\leftarrow x_0\oplus x_2$ & $\{x_0\}$ & $\{x_1\}$ & $\{x_0\}$ & $\{x_1,x_2\}$ & load $x_2$ & 1 \\
3 & $x_1\leftarrow x_1\oplus x_3$ & $\{x_0\}$ & $\{x_1,x_2\}$ & $\{x_0,x_1\}$ & $\{x_3\}$ & load $x_3$ (drop $x_2$) & 1 \\
4 & $x_0\leftarrow x_0\oplus x_2$ & $\{x_0,x_1\}$ & $\{x_3\}$ & $\{x_0,x_1\}$ & $\{x_2\}$ & load $x_2$ (drop $x_3$)  & 1 \\
5 & $x_1\leftarrow x_1\oplus x_2$ & $\{x_0,x_1\}$ & $\{x_2\}$ & $\{x_0,x_1\}$ & $\{x_2\}$ & - & 0 \\
\bottomrule
\end{tabular}
}
\end{table}

\begin{figure}[!h]
\centering
\includegraphics[width=0.8\linewidth]{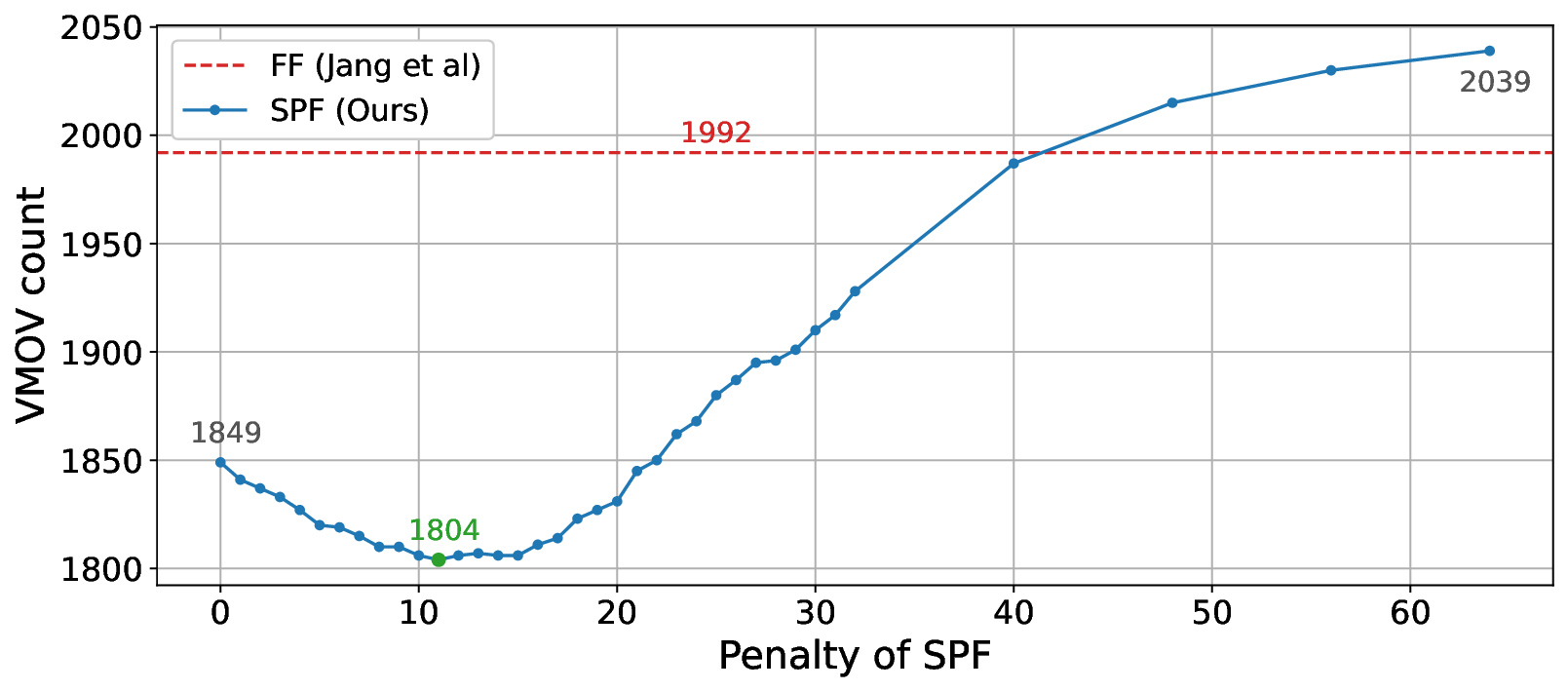}
\caption{VMOV count as a function of the SPF penalty $p$ over all $v_i$ for \HQC{5}.}
\label{fig:SPF-penalty}
\end{figure}

To assess the effect of the penalty $p$, we measure the VMOV count for the twelve butterfly functions ($v_{17},\dots,v_{28}$) of \HQC{5} while varying $p$ and keeping the XOR sequences of the implementation of Jang et al.~\cite{jlk25} fixed. As Figure~\ref{fig:SPF-penalty} shows, the count is $1{,}849$ at $p=0$, decreases as $p$ increases, reaches its minimum of $1{,}804$ at $p=11$, and then rises again, forming a U shape. This reflects a trade-off between avoiding dirty spills and incurring additional clean reloads: a small $p$ makes SPF behave much like FF and underprotect dirty variables, whereas a large $p$ evicts clean variables that will soon be reused and thus increases load VMOVs. We therefore use $p=11$, which yields the lowest VMOV count, as the default penalty for SPF.

\subsubsection{Reducing VMOV Costs via XOR Reordering}
\label{sec:vmov-framework}

The implementation cost model of \Sref{sec:vmov-cost-model} deterministically computes the VMOV count of an XOR sequence $L$ under a fixed register-allocation policy. In this subsection, among the XOR sequences that preserve the resulting matrix $M$, we search for a new order that reduces the VMOV cost.

\paragraph{Commutativity and the Search Space}
\label{sec:vmov-theory-commute}

The VMOV count depends not only on the register-allocation policy but also on the order of the type-3 operations. We therefore rearrange only the order of operations while preserving the multiset of XOR operations and the resulting matrix $M$. Property~\ref{prop:xiang3} gives three cases in which two type-3 elementary matrices commute: Case~1, where the two operations use disjoint indices, and Cases~2 and~3, where they share the same source and the same destination, respectively.

\begin{property}[{Xiang et al.~\cite{xiang2020linear}, Property~3}]
\label{prop:xiang3}
For distinct integers $i, j, k, l$, the following identities hold.
\begin{align*}
   \mathrm{Case}~1.\;& E(k+l)\,E(i+j) = E(i+j)\,E(k+l),\\
   \mathrm{Case}~2.\;& E(i+j)\,E(k+j) = E(k+j)\,E(i+j),\\
   \mathrm{Case}~3.\;& E(i+j)\,E(i+k) = E(i+k)\,E(i+j).
\end{align*}
\end{property}

Two operations are said to \emph{commute} if they satisfy one of the three identities of Property~\ref{prop:xiang3}. A non-commuting pair must retain its original relative order, and for each such pair, we add an edge from the earlier operation to the later one, forming a dependency graph $D$. A commuting pair, by contrast, preserves the resulting matrix $M$ under a swap of its order and is therefore a candidate for reordering to reduce the VMOV cost.

\Tref{tab:xor-reordering-example} is a simple example showing that reordering can reduce the VMOV count, using the same XOR sequence as the FF and SPF examples above with $K=3$ GP registers and the SPF policy. In the original order $L$, $x_2$ is loaded at $o_2$ but evicted at $o_3$ to make room for $x_3$, and must therefore be reloaded at $o_4$. Since $o_2 = E(0{+}2)$ and $o_3 = E(1{+}3)$ use disjoint indices, they commute by Case~1 of Property~\ref{prop:xiang3}. In the reordered order $L'$, $o_3$ runs first and loads $x_3$ while a register is still free, and $x_2$ is then loaded at $o_2$ and stays resident through $o_4$ and $o_5$, so the reload disappears. Both orders implement the same $M$, yet $L'$ uses one fewer VMOV, showing that reordering alone---while preserving the resulting matrix $M$---can change register pressure and thereby lower the VMOV cost.

\begin{table}[!h]
\centering
\caption{Example of VMOV reduction by commuting two XOR operations.}
\label{tab:xor-reordering-example}
\small
\resizebox{\linewidth}{!}{%
\begin{tabular}{c|l|l|c|l|l|c}
\toprule
$t$ & \multicolumn{3}{c|}{Original order $L$} & \multicolumn{3}{c}{Reordered order $L'$} \\
\cmidrule(lr){2-4}\cmidrule(lr){5-7}
 & XOR operation & Required \texttt{VMOV} & \#\texttt{VMOV}
 & XOR operation & Required \texttt{VMOV} & \#\texttt{VMOV} \\
\midrule
1 & $o_1: x_0\leftarrow x_0\oplus x_1$ & load $x_0$, load $x_1$ & 2
  & $o_1: x_0\leftarrow x_0\oplus x_1$ & load $x_0$, load $x_1$ & 2 \\
2 & $o_2: x_0\leftarrow x_0\oplus x_2$ & load $x_2$ & 1
  & $\mathbf{o_3}: x_1\leftarrow x_1\oplus x_3$ & load $x_3$ & 1 \\
3 & $o_3: x_1\leftarrow x_1\oplus x_3$ & load $x_3$ (drop $x_2$) & 1
  & $\mathbf{o_2}: x_0\leftarrow x_0\oplus x_2$ & load $x_2$ (drop $x_3$) & 1 \\
4 & $o_4: x_0\leftarrow x_0\oplus x_2$ & load $x_2$ (drop $x_3$) & 1
  & $o_4: x_0\leftarrow x_0\oplus x_2$ & none & 0 \\
5 & $o_5: x_1\leftarrow x_1\oplus x_2$ & none & 0
  & $o_5: x_1\leftarrow x_1\oplus x_2$ & none & 0 \\
\midrule
\multicolumn{3}{r|}{Total} & 5
  & \multicolumn{2}{r|}{Total} & 4 \\
\bottomrule
\end{tabular}%
}
\end{table}

\paragraph{Reordering via Simulated Annealing}
\label{sec:vmov-sa}

The number of reordering candidates that preserve the resulting matrix $M$ varies greatly with the structure of the dependencies, so an exhaustive search over all possible orders is impractical. Moreover, how much a local change in operation order alters the VMOV count must itself be recomputed by the SPF-based forward simulation. For such a problem---with a large search space and a cost that must be evaluated directly for each candidate---stochastic local search, which repeatedly generates and evaluates candidates, is well suited. We therefore use simulated annealing~\cite{kirkpatrick1983optimization} to search for XOR sequences with a small VMOV count. A plain greedy search accepts only cost-decreasing moves and is thus prone to getting stuck in local minima, whereas simulated annealing also accepts cost-increasing moves probabilistically, allowing broad exploration early on and convergence around improved candidates later.

We search over the orders of the XOR sequence using the SPF-based VMOV count as the objective. A neighboring candidate is generated by an insertion move that picks one operation and moves it to another position, allowed only when the moved operation commutes with every operation it passes. Since such a move is a composition of swaps of adjacent commuting pairs, the resulting XOR sequence preserves the resulting matrix $M$. With probability $1/2$ a move swaps two adjacent operations, and otherwise it relocates an operation to a position within distance $w$; the former fine-tunes the cost, while the latter provides the long-range moves needed to escape local minima. Each new candidate is accepted according to the Metropolis rule: a move is always accepted if it does not increase the cost, and a move that increases the cost by $\Delta$ is accepted with probability $e^{-\Delta/t}$. The temperature $t$ is lowered geometrically from $t_0$ to $t_1$.

\paragraph{Experimental Results}
\label{sec:vmov-results}

\Tref{tab:vmov-merged} compares, for each parameter, our implementation against a prior one that uses the same XOR sequences, so the EOR count of each function is identical and only the VMOV count differs. The reference is the implementation of Jang et al.~\cite{jlk25} for \HQC{5} and of Chen et al.~\cite{TCHES:CCC26} for \HQC{3}. For \HQC{1}, the new evaluation set we choose for the FAFFT (\Sref{sec:polymul-basis}) induces XOR sequences that have no prior implementation to compare against, so we take as the reference these same sequences allocated with the farthest-first policy of Lee et al.~\cite{MyeonghoonIOT} (marked with an asterisk). The number of these fixed-constant multiplication functions is $10$, $11$, and $12$ for \HQC{1}, \HQC{3}, and \HQC{5}, respectively. Keeping the XOR sequences fixed, our optimization reduces only the VMOV count by changing the operation order and register allocation, lowering the total VMOV count by $40.6\%$ (from $1{,}866$ to $1{,}109$) for \HQC{1}, $48.1\%$ (from $2{,}396$ to $1{,}243$) for \HQC{3}, and $37.6\%$ (from $1{,}992$ to $1{,}244$) for \HQC{5}.

\begin{table*}[!h]
\centering
\caption{EOR and VMOV counts per function under XOR reordering.}
\label{tab:vmov-merged}
\resizebox{\textwidth}{!}{%
\begin{tabular}{llrrrrrrrrrrrrr}
\toprule
\multicolumn{2}{c}{$\mathrm{inv}_i$}
& $\mathrm{inv}_{1}$ & $\mathrm{inv}_{2}$ & $\mathrm{inv}_{3}$ & $\mathrm{inv}_{4}$
& $\mathrm{inv}_{5}$ & $\mathrm{inv}_{6}$ & $\mathrm{inv}_{7}$ & $\mathrm{inv}_{8}$
& $\mathrm{inv}_{9}$ & $\mathrm{inv}_{10}$ & $\mathrm{inv}_{11}$ & $\mathrm{inv}_{12}$
& Total \\

\midrule
\midrule
\multicolumn{15}{c}{\textbf{\HQC{1}}} \\
\midrule
\multicolumn{2}{l}{\texttt{EOR}}  & 183 & 187 & 163 & 184 & 189 & 192 & 186 & 187 & 186 & 183 & $-$ & $-$ & 1{,}840 \\
\midrule
\multirow{4}{*}{\texttt{VMOV}}
& \cite{MyeonghoonIOT}$^{*}$ & 197 & 185 & 163 & 181 & 185 & 209 & 197 & 199 & 173 & 177 & $-$ & $-$ & 1{,}866 \\
& Ours & 117 & 110 & 101 & 104 & 113 & 121 & 110 & 112 & 113 & 108 & $-$ & $-$ & \textbf{1{,}109} \\
\cmidrule(lr){2-15}
& $\Delta$ & $-80$ & $-75$ & $-62$ & $-77$ & $-72$ & $-88$ & $-87$ & $-87$ & $-60$ & $-69$ & $-$ & $-$ & $\mathbf{-757}$ \\

\midrule
\midrule
\multicolumn{15}{c}{\textbf{\HQC{3}}} \\
\midrule
\multicolumn{2}{l}{\texttt{EOR}}
& 167 & 182 & 180 & 183 & 186 & 192 & 193 & 182 & 187 & 185 & 189 & $-$ & 2{,}026 \\
\midrule
\multirow{4}{*}{\texttt{VMOV}}
& \cite{TCHES:CCC26}
& 194 & 214 & 216 & 200 & 238 & 236 & 226 & 204 & 220 & 224 & 224 & $-$ & 2{,}396 \\
& Ours
& 105 & 114 & 114 & 110 & 115 & 117 & 114 & 109 & 116 & 109 & 120 & $-$ & \textbf{1{,}243} \\
\cmidrule(lr){2-15}
& $\Delta$
& $-89$ & $-100$ & $-102$ & $-90$ & $-123$ & $-119$ & $-112$ & $-95$ & $-104$ & $-115$ & $-104$ & $-$ & $\mathbf{-1{,}153}$ \\

\midrule
\midrule
\multicolumn{15}{c}{\textbf{\HQC{5}}} \\
\midrule
\multicolumn{2}{l}{\texttt{EOR}}
& 159 & 153 & 166 & 172 & 166 & 160 & 163 & 180 & 180 & 179 & 173 & 183 & 2{,}034 \\
\midrule
\multirow{4}{*}{\texttt{VMOV}}
& \cite{jlk25}
& 153 & 139 & 161 & 167 & 165 & 159 & 165 & 171 & 171 & 187 & 181 & 173 & 1{,}992 \\
& Ours
& 100 & 88 & 105 & 110 & 103 & 98 & 97 & 104 & 110 & 110 & 111 & 108 & \textbf{1{,}244} \\
\cmidrule(lr){2-15}
& $\Delta$
& $-53$ & $-51$ & $-56$ & $-57$ & $-62$ & $-61$ & $-68$ & $-67$ & $-61$ & $-77$ & $-70$ & $-65$ & $\mathbf{-748}$ \\
\bottomrule
\end{tabular}%
}
\par\smallskip
{\footnotesize\raggedright $^{*}$ Our \HQC{1} XOR sequence has no prior implementation; the reference is that same sequence allocated with the farthest-first policy of \Sref{sec:vmov-theory-count},
following the implementation of Lee et al.~\cite{MyeonghoonIOT}.\par}
\end{table*}

\subsection{Multiplication Performance}
\label{sec:polymul-eval}

We benchmark the polynomial multiplication in isolation on the NUCLEO-L4R5ZI at $20$\,MHz, comparing our implementation against those of Chen et al.~\cite{TCHES:CCC26} and Jang et al.~\cite{jlk25}. These figures reflect both the multiplication construction of \Sref{sec:polymul-combine} and the VMOV minimization of \Sref{sec:vmov-min}, so the comparison captures the joint effect of the two. \Tref{tab:polymul-cycles} reports the results.

\begin{table}[!h]
\centering
\caption{Clock cycles of the polynomial multiplication.}
\label{tab:polymul-cycles}
\small
\begin{tabular}{lrrrr}
\toprule
Parameter & \cite{TCHES:CCC26} & \cite{jlk25} & Ours & Reduction \\
\midrule
\HQC{1} & $1{,}753{,}860$ & $1{,}819{,}624$ & $1{,}574{,}380$ & $-10.2\%$ \\
\HQC{3} & $4{,}836{,}147$ & $4{,}441{,}928$ & $3{,}984{,}093$ & $-10.3\%$ \\
\HQC{5} & $6{,}290{,}041$ & $6{,}103{,}141$ & $6{,}038{,}662$ & $-1.1\%$ \\
\bottomrule
\end{tabular}
\end{table}

Our multiplication is faster than both prior implementations; relative to the faster of the two prior implementations, the multiplication cycles drop by $10.2\%$ for \HQC{1}, $10.3\%$ for \HQC{3}, and $1.1\%$ for \HQC{5}. The improvement is largest for \HQC{3}, where the CRT decomposition with truncated basis conversion removes most of the forward-transform overhead, and smallest for \HQC{5}, whose multiplication is a pure double-size FAFFT with no residue ring and therefore benefits least from the construction of \Sref{sec:polymul-combine}. 

\section{Fixed-Weight Sampling Optimization}
\label{sec:vectwrite}

HQC samples the fixed-weight sparse vectors $\mathbf{x}, \mathbf{y}$ from a secret seed and $\mathbf{r}_1, \mathbf{r}_2, \mathbf{e}$ as encryption randomness throughout key generation, encapsulation, and decapsulation. This sampling proceeds in two steps. First, \texttt{vect\_generate\_random\_support1/2} extracts the support, the $\omega$ distinct positions, from an XOF. Then \texttt{vect\_write\_support\_to\_vector} (WSV) expands that support into a length-$n$ bit vector. This section proposes an optimization that performs the latter scatter operation in constant time on the Cortex-M4.

\subsection{Constant-Time Support Expansion}
\label{sec:vectwrite-baseline}

The WSV function is called after each \texttt{vect\_generate\_random\_support1/2} to expand a fixed-weight vector's support, and is performed throughout the KEM operations. Concretely, it is called twice for $\mathbf{x}, \mathbf{y}$ during key generation, three times for $\mathbf{r}_1, \mathbf{r}_2, \mathbf{e}$ during encapsulation, and four times during decapsulation, the latter including the re-encryption for the FO transform in addition to recovering the secret vector $\mathbf{y}$.

The support produced by \texttt{vect\_generate\_random\_support1/2} is the set of indices where the bit $1$ appears in the secret vector. Hence a naive expansion that references each position $p_j$ directly as a memory address index, as in $\mathbf{v}[\,p_j \gg 6\,] \mathrel{|}= 1 \ll (p_j \,\&\, 63)$, induces a secret-dependent memory access pattern and is exposed to timing and cache side-channel attacks. To prevent this, the official HQC implementation first converts the support into word indices and bit masks, and then scans over every combination of the number of output words $n_{64} = \lceil n/64 \rceil$ and the support weight $\omega$ without branching. The implementations of~\cite{TCHES:CCC26, jlk25} also adopt this expansion scheme unchanged. Since the preprocessing runs only $\omega$ times, the expansion overhead is concentrated mostly in the inner double loop that iterates $n_{64} \times \omega$ times.

The number of conditional bit accumulations inside the double loop, per single expansion and based on the Hamming weight $\omega_r = \omega_e$ of $\mathbf{r}_1, \mathbf{r}_2, \mathbf{e}$, is $277 \times 75 = 20{,}775$ for \HQC{1}, $561 \times 114 = 63{,}954$ for \HQC{3}, and $901 \times 149 = 134{,}249$ for \HQC{5}. As a result, for the \HQC{1} performance measured on the NUCLEO-L4R5ZI board, this expansion accounts for about $29.4\%$ of the key-generation cycles, $28.2\%$ of encapsulation, and $21.6\%$ of decapsulation relative to the implementation of~\cite{TCHES:CCC26}, as shown in \Tref{tab:vw-baseline}.

\begin{table}[h]
\centering
\caption{Proportion of the WSV operation in the total cycles for \HQC{1}, measured on NUCLEO-L4R5ZI relative to the implementation of~\cite{TCHES:CCC26}.}
\label{tab:vw-baseline}
\begin{tabular}{lrrr}
\toprule
Operation & Expansion cycles & Full operation & Proportion \\
\midrule
Key generation & $859{,}628$  & $2{,}924{,}928$ & $29.4\%$ \\
Encapsulation & $1{,}461{,}071$ & $5{,}181{,}432$ & $28.2\%$ \\
Decapsulation & $1{,}889{,}926$ & $8{,}738{,}801$ & $21.6\%$ \\
\bottomrule
\end{tabular}
\end{table}

Prior implementations first decompose each support element into a word index $\mathit{idx}_j = p_j \gg 6$ and a bit mask $\mathit{bit}_j = 1 \ll (p_j \,\&\, 63)$, and then, for each output word index $i$, scan over all $j$ and accumulate using the following portable branchless mask technique:
\begin{equation}
   \mathit{mask} = -\bigl(1 \oplus ((t \mathbin{|} -t) \gg 31)\bigr), \quad t = i - \mathit{idx}_j, \qquad
   \mathbf{v}[i] \mathrel{|}= \mathit{bit}_j \,\&\, \mathit{mask}.
   \label{eq:vw-baseline}
\end{equation}
This mask computation guarantees constant-time execution, but generating the mask and applying it at every comparison step inside the loop requires a total of eight integer instructions.

\subsection{Optimization Method}
\label{sec:vectwrite-method}

We replace the mask computation of~\eqref{eq:vw-baseline} with the predicated execution feature of the ARMv7E-M architecture, and process the output words in groups of four to reduce the per-word comparison and load costs. Because 4-way loop unrolling requires many registers, we address it by designing the core loop of the expansion as a dedicated assembly routine.

The pre-decomposition that converts each support position $p_j$ into a word index $\mathit{idx}_j = p_j \gg 6$ and a bit mask $\mathit{bit}_j = 1 \ll (p_j \,\&\, 63)$ is, as before, computed once in a preprocessing step and reused across the inner loop. The complete optimized inner loop of the proposed expansion is given in \Aref{alg:vw}. Below we describe it from three perspectives in order, namely predicated scatter, 4-way unrolling, and register allocation.

\begin{algorithm}[!ht]
\caption{Predicated 4-way scatter inner loop.}
\label{alg:vw}
\begin{lstlisting}[language=armasm]
.Lloop:
    ldmia r12!, {r0, r1, lr}   @ idx, bit_lo, bit_hi  (ptr += 12)
    subs  r0, r3, r0           @ d = i - idx_j
    itt   eq                   @ if (d == 0)   : word i
    orreq r4, r4, r1
    orreq r5, r5, lr
    cmn   r0, #1               
    itt   eq                   @ if (d+1 == 0) : word i+1
    orreq r6, r6, r1
    orreq r7, r7, lr
    cmn   r0, #2              
    itt   eq                   @ if (d+2 == 0) : word i+2
    orreq r8, r8, r1
    orreq r9, r9, lr
    cmn   r0, #3              
    itt   eq                   @ if (d+3 == 0) : word i+3
    orreq r10, r10, r1
    orreq r11, r11, lr
    cmp   r12, r2              @ loop until r12 reaches r2 (end ptr)
    bne   .Lloop
\end{lstlisting}
\end{algorithm}

\paragraph{Predicated scatter.} The key idea is to handle the conditional accumulation, which ORs the bit mask $\mathit{bit}_j$ into the output vector $\mathbf{v}[i]$ only when $i = \mathit{idx}_j$, directly through the conditional instruction \texttt{orreq}, without going through a mask-computation step. Once the \texttt{subs} instruction computes $i - \mathit{idx}_j$ and sets the $Z$ flag, the following \texttt{orreq} executes only at the moment $i = \mathit{idx}_j$ and leaves the value unchanged otherwise. Because this predicated execution introduces no data-dependent branch and each \texttt{orreq} costs the same whether or not its condition holds, the instruction flow and cycle count are fixed independently of the concrete value of $\mathit{idx}_j$, guaranteeing constant time.

\paragraph{4-way unrolling.} Using only a single \texttt{subs} comparison, we update the four output words $i, i{+}1, i{+}2, i{+}3$ at once. We keep $d = i - \mathit{idx}_j$, the result of \texttt{subs}, and execute \texttt{cmn d, \#1}, \texttt{cmn d, \#2}, and \texttt{cmn d, \#3} to check whether $\mathit{idx}_j$ equals $i{+}1, i{+}2, i{+}3$ without any additional subtraction. This reduces the comparison to one per output word, and the load cost for the upper and lower 32-bit halves of $\mathit{idx}_j$ and $\mathit{bit}_j$, previously incurred three times per support element, is spread evenly across the four words down to $0.75$ per word. Meanwhile, since each 64-bit word is accumulated in \texttt{hi}/\texttt{lo} halves to fit the Cortex-M4's 32-bit register width, processing four words uses eight registers as accumulators, \texttt{r4}--\texttt{r11} in \Aref{alg:vw}. For HQC parameter sets in which $n_{64}$ is not a multiple of four, the words remaining after the loop, at most three, are finished off by a scalar tail loop functionally identical to~\eqref{eq:vw-baseline}.

\paragraph{Register allocation.} The eight accumulators, the two upper/lower halves of the loaded bit mask, the temporary result $d$, and the streaming pointer to the packed support table together with the end pointer that marks loop termination must all stay resident in registers. The number of values that must be simultaneously live thus reaches thirteen, whereas the Cortex-M4 provides only fourteen general-purpose registers. At the \texttt{-O3} level the compiler cannot allocate these together with the registers needed for control operations such as the loop counter and memory base-address computation, causing a bottleneck. To resolve this limitation, we implement the core expansion loop of WSV as a hand-written dedicated assembly routine and allocate all fourteen available general-purpose registers of the Cortex-M4 directly. The concrete techniques are as follows. First, we design a packed table that interleaves each support element as $\{\mathit{idx}, \mathit{bit}_{\mathrm{lo}}, \mathit{bit}_{\mathrm{hi}}\}$. A single memory pointer and one \texttt{ldmia} instruction then handle the data load, removing the overhead of multiple pointers occupying registers. Second, temporaries that stay fixed within the loop body or are rarely used, such as the start address of the output buffer, are temporarily held on the stack to free up register space.

\paragraph{Constant-time guarantee.} In addition to the branchless predication above, the loop bounds $n_{64}$ and $\omega$ depend only on public system parameters, and the memory access is data-independent, loading the $\mathit{idx}_j$ and $\mathit{bit}_j$ data and writing $\mathbf{v}$ in a fixed sequential order. Thus the total execution time is independent of the input support.

\subsection{Results}
\label{sec:vectwrite-results}

Measuring the expansion on the NUCLEO-L4R5ZI board under the same conditions as \Tref{tab:vw-baseline}, the proposed predicated 4-way expansion reduces the expansion cycles by about $73.5\%$ to $73.7\%$ over the prior implementation, as shown in \Tref{tab:vw-result}. In detail, for \HQC{1} it reduces key generation from $859{,}628$ to $227{,}379$ cycles, encapsulation from $1{,}461{,}071$ to $384{,}121$ cycles, and decapsulation from $1{,}889{,}926$ to $496{,}851$ cycles, corresponding to speedups of about $3.78\times$, $3.80\times$, and $3.80\times$, respectively. This stems from replacing the prior eight mask instructions with one \texttt{subs}/\texttt{cmn} and two \texttt{orreq} conditional operations per word, and from loading each support element once with \texttt{ldmia} and reusing it across the four words. At the instruction level as well, the prior innermost loop takes about $22$ cycles per support-element and output-word pair including the loads and mask computation, whereas the proposed method processes four words at once and lowers this to about $6$ cycles per word. A KEM-level evaluation combining this method with the other optimizations is presented in \Sref{sec:evaluation}.

\begin{table}[h]
\centering
\caption{WSV cycle comparison for \HQC{1}.}
\label{tab:vw-result}
\begin{tabular}{lrrr}
\toprule
Operation & \cite{TCHES:CCC26} & Ours & Reduction \\
\midrule
keypair & $859{,}628$     & $227{,}379$ & $73.5\%$ \\
encaps  & $1{,}461{,}071$ & $384{,}121$ & $73.7\%$ \\
decaps  & $1{,}889{,}926$ & $496{,}851$ & $73.7\%$ \\
\bottomrule
\end{tabular}
\end{table}

\section{Caching Strategies}
\label{sec:caching}

Across repeated invocations under the same key, HQC's encapsulation and decapsulation recompute the same public values from scratch on every call. The public polynomial $\mathbf{h}$ is regenerated from $\mathit{seed}_{\mathsf{ek}}$ via SHAKE at every operation, and the FAFFT forward transforms of $\mathbf{h}$ and $\mathbf{s}$ used in multiplication, together with the public-key hash, are recomputed each time. In a setting that repeats many operations under one fixed key, this recomputation is duplicated between calls. Chen et al.~\cite{TCHES:CCPY26} proposed a memoization technique that stores and reuses such transformed keys for HQC in a SIMD setting. This section ports that idea to the Cortex-M4 and proposes an optional caching that computes these public values once, stores them in a single cache identified by the public key, and reuses them on subsequent operations. The cache resides entirely in local RAM and leaves the transmitted public-key size unchanged, so as a time--space tradeoff its use can be decided at build time.

\subsection{Applicability and Scenarios}
\label{sec:caching-scenario}

The benefit of caching is realized when operations repeat under the same key, and the scenario in which this repetition is structurally guaranteed is decapsulation. A device that many peers connect to using the device's fixed public key, such as a Cortex-M4-class access-control reader, fare gate, or payment terminal, repeats decapsulation with its own fixed key on every connection. On such a device the cache is filled on the first call and reused by all subsequent decapsulation calls, and because the key belongs to the device itself the validity of reuse is structurally guaranteed. In contrast, encapsulation caching is confined to specific architectures where operations repeat for the same recipient's public key, and in ordinary session-based communication encapsulation runs only once per session, so cache reuse is hard to expect. Key generation is a one-time operation and is unaffected by caching. Under these scenarios, the measurements in this section assume a steady state in which the same public key is continuously reused.

\subsection{Caching Targets and Mechanism}
\label{sec:caching-target}

As explained in \Sref{sec:preliminaries}, $\mathbf{h}$ is a public polynomial generated deterministically from $\mathit{seed}_{\mathsf{ek}}$. Both encapsulation and decapsulation perform the multiplications $\mathbf{u} = \mathbf{r}_1 + \mathbf{h}\cdot\mathbf{r}_2$ and $\mathbf{s}\cdot\mathbf{r}_2$ when forming the ciphertext, and decapsulation recomputes them in the re-encryption of the Fujisaki--Okamoto (FO) transform. The forward transforms of $\mathbf{h}$ and $\mathbf{s}$ are therefore needed in both operations. In addition, both operations compute the public-key hash $\mathtt{H}(\mathsf{ek})$, where $\mathsf{ek}$ is the KEM encapsulation key, to derive the shared secret and the encryption seed. This hash and the two forward transforms all depend only on the fixed public key, so they recur identically across operations.

The cached public values are the following four, and only the form of the transform varies with the parameter set.
\begin{itemize}
\item \textbf{Public key $\mathsf{ek}$}: used as the key that identifies the cache.
\item \textbf{Public-key hash $\mathtt{H}(\mathsf{ek})$}: used as the hash input for deriving the shared secret and the encryption seed in encapsulation and decapsulation.
\item \textbf{Transform of $\mathbf{h}$}: all three parameter sets store the forward transform $\mathsf{FAFFT}(\mathbf{h})$. \HQC{1} and \HQC{3} additionally store the low-order coefficients of $\mathbf{h}$ needed for the auxiliary CRT multiplication.
\item \textbf{Transform of $\mathbf{s}$}: all three parameter sets store the forward transform $\mathsf{FAFFT}(\mathbf{s})$. Since $\mathbf{s}$ is carried directly in the public key, its low-order coefficients are not stored separately.
\end{itemize}

All of these items are already externally exposed public information, so caching and reusing them does not compromise security, whereas caching secret-dependent data would risk side-channel leakage. They are simply public values regenerated and transformed from the seed on each call, now kept resident in RAM, trading memory for speed.

The core of the proposed structure is that the $\mathtt{hash\_h}$ function performing the public-key hashing is designed as the single decision point of the cache. Since both encapsulation and decapsulation compute this hash in common before the polynomial multiplication, at that step the incoming public key $\mathsf{ek}'$ is compared against the $\mathsf{ek}$ stored in the cache by a constant-time comparison over the full data length. On a cache hit, where the entry's validity flag is set ($\mathit{valid}=1$) and the two keys match, the stored hash result is returned and the existing transform data is judged valid. On a cache miss, where they do not match, the new public key and hash value are written to the cache and the internal polynomial transforms are marked as an empty, not-yet-updated state. When the first subsequent multiplication then detects that the transform cache is empty, the polynomial $\mathbf{h}$ is regenerated from the public-key seed and forward-transformed, $\mathbf{s}$ is also forward-transformed, the cache is filled, and the valid bit is set. While the valid bit is set, the multiplication uses the cached transforms of $\mathbf{h}$ and $\mathbf{s}$ by direct reference without any additional condition check.

The proposed caching mechanism preserves the constant-time property of the entire operation. The public-key comparison accumulates bitwise results over the full key length without any conditional, and no secret-dependent branch or memory access occurs during cache lookup or data loading. We confirmed that the code with caching enabled passes the official KAT vectors. The speed and memory effects of the caching are evaluated in \Sref{sec:evaluation}.

\section{Evaluation}
\label{sec:evaluation}


This section evaluates the speed performance and memory overhead of the proposed HQC optimizations on a Cortex-M4, and compares them with the two prior state-of-the-art implementations, \cite{TCHES:CCC26} and \cite{jlk25}. All measurements were taken on an STM32 NUCLEO-L4R5ZI development board, which carries an ARM Cortex-M4 core with 2\,MB of Flash and 640\,KB of SRAM, using the pqm4 benchmarking framework~\cite{PQM4}. The CPU clock was set to 20\,MHz so that measurements ran with zero wait states. To ensure a fair comparison with the prior implementations, we used the same arm-none-eabi-gcc 10.3.1 compiler with the \texttt{-O3} option.

\subsection{Speed Performance}
\Tref{tab:eval-speed} compares the cycle counts of the HQC with our optimizations against the prior state-of-the-art implementations. Here the proposed method is the optimization without caching, incorporating the FFT butterfly register-scheduling optimization of \Sref{sec:vmov} and the 4-way-unrolled support-expansion method of \Sref{sec:vectwrite}.

Relative to the fastest prior implementation for each operation, the proposed method reduces cycles by $27.7\%$ to $33.1\%$ for Keypair, $27.6\%$ to $34.6\%$ for Encaps, and $22.0\%$ to $29.8\%$ for Decaps. The additional reduction from the optional caching is treated separately in \Sref{sec:eval-caching}.

\begin{table}[h!]
\centering
\caption{Cycle counts of HQC implementations on Cortex-M4.}
\label{tab:eval-speed}
\small
\begin{tabular}{llrrrr}
\toprule
Parameter & Operation & \cite{TCHES:CCC26} & \cite{jlk25} & Ours & Reduction \\
\midrule
\multirow{3}{*}{\HQC{1}}
 & Keypair & $2{,}924{,}928$ & $2{,}993{,}695$ & $2{,}113{,}726$ & $-27.7\%$ \\
 & Encaps  & $5{,}181{,}432$ & $5{,}798{,}993$ & $3{,}753{,}019$ & $-27.6\%$ \\
 & Decaps  & $8{,}738{,}801$ & $9{,}959{,}877$ & $6{,}815{,}280$ & $-22.0\%$ \\
\midrule
\multirow{3}{*}{\HQC{3}}
 & Keypair & $8{,}053{,}779$ & $7{,}659{,}978$ & $5{,}262{,}127$ & $-31.3\%$ \\
 & Encaps  & $14{,}425{,}876$ & $14{,}667{,}541$ & $9{,}428{,}668$ & $-34.6\%$ \\
 & Decaps  & $22{,}872{,}448$ & $23{,}352{,}058$ & $16{,}053{,}394$ & $-29.8\%$ \\
\midrule
\multirow{3}{*}{\HQC{5}}
 & Keypair & $12{,}701{,}978$ & $12{,}513{,}535$ & $8{,}369{,}531$ & $-33.1\%$ \\
 & Encaps  & $21{,}890{,}091$ & $23{,}735{,}473$ & $14{,}510{,}838$ & $-33.7\%$ \\
 & Decaps  & $34{,}769{,}358$ & $38{,}191{,}910$ & $25{,}097{,}439$ & $-27.8\%$ \\
\bottomrule
\end{tabular}
\end{table}

\subsection{Evaluation of the Caching Strategy}
\label{sec:eval-caching}

In addition to the algorithmic optimizations described above, to isolate the standalone benefit of the caching proposed in \Sref{sec:caching}, we compare the cycle counts of HQC with and without caching under the repeated-same-key scenario. The memory effect of caching is discussed together with the overall memory usage in \Sref{sec:eval-memory}.

As shown in \Tref{tab:eval-caching-speed}, in the steady state where the same public key is reused, caching further improves performance. Encapsulation cycles decrease by about $30.3\%$, $26.0\%$, and $32.7\%$ for \HQC{1}, \HQC{3}, and \HQC{5} relative to the case without caching, and decapsulation by about $16.7\%$, $15.3\%$, and $18.9\%$. The absolute cycle savings of the two operations are similar because caching removes the common fixed cost of regenerating the public polynomial $\mathbf{h}$, forward-transforming $\mathbf{h}$ and $\mathbf{s}$, and computing the public-key hash $\mathtt{H}(\mathsf{ek})$. Key generation is not a caching target, so only encapsulation and decapsulation are compared.

\begin{table}[h!]
\centering
\caption{Cycle counts of the proposed method with and without caching.}
\label{tab:eval-caching-speed}
\small
\begin{tabular}{llrrr}
\toprule
Parameter & Operation & No caching & Caching & Reduction \\
\midrule
\multirow{2}{*}{\HQC{1}}
 & Encaps  & $3{,}753{,}019$  & $2{,}616{,}820$  & $-30.3\%$ \\
 & Decaps  & $6{,}815{,}280$  & $5{,}679{,}147$  & $-16.7\%$ \\
\midrule
\multirow{2}{*}{\HQC{3}}
 & Encaps  & $9{,}428{,}668$  & $6{,}976{,}592$  & $-26.0\%$ \\
 & Decaps  & $16{,}053{,}394$ & $13{,}601{,}480$ & $-15.3\%$ \\
\midrule
\multirow{2}{*}{\HQC{5}}
 & Encaps  & $14{,}510{,}838$ & $9{,}769{,}663$  & $-32.7\%$ \\
 & Decaps  & $25{,}097{,}439$ & $20{,}356{,}262$ & $-18.9\%$ \\
\bottomrule
\end{tabular}
\end{table}

\subsection{Memory Usage}
\label{sec:eval-memory}
\Tref{tab:eval-memory} reports the Flash code size, static RAM, and dynamic stack usage of the proposed method, with and without caching. The proposed method builds on the FAFFT--CRT multiplication of \cite{jlk25} and additionally adopts the RS/RM decoder of \cite{TCHES:CCC26} to accelerate the decoding in decapsulation.

The RAM usage of the proposed method stems from the FAFFT--CRT multiplication path it shares with \cite{jlk25}. For \HQC{1} and \HQC{3}, which use CRT, the RAM of the proposed method is 14,344 and 37,504 bytes, larger than the implementation of \cite{TCHES:CCC26} but equal to or even smaller than that of \cite{jlk25}, which shares the same multiplication. This difference stems from the choice of CRT residue multiplication. \HQC{3} uses the radix-16 residue multiplication of \cite{jlk25}, so its RAM matches \cite{jlk25}, whereas \HQC{1} adopts the radix-16 residue multiplication of \cite{TCHES:CCC26} and is 2,976 bytes smaller than \cite{jlk25}. For \HQC{5}, which does not use CRT, all three implementations share the same RAM of 2,824 bytes.

The Flash increase comes from the RS/RM decoder of \cite{TCHES:CCC26} adopted to accelerate decapsulation. This decoder unrolls the radix-16 matrix--vector products to gain speed at the cost of larger code, so relative to \cite{jlk25} the Flash of the proposed method is larger by 33,736, 35,168, and 162,912 bytes for \HQC{1}, \HQC{3}, and \HQC{5}. The increase is most pronounced for \HQC{5} because \HQC{5} uses a larger RS code, which enlarges the unrolled matrix--vector products. Compared with the implementation of \cite{TCHES:CCC26}, which uses the same decoder, the Flash of the proposed method is instead 6,208 bytes smaller for \HQC{5}. They fit comfortably within the 2\,MB Flash and 640\,KB SRAM of the target board, and the stack usage stays within a range similar to the two prior implementations.

Caching trades this speedup for additional static memory. As the Ours (caching) rows of \Tref{tab:eval-memory} show, caching increases RAM by $10{,}792$, $21{,}704$, and $40{,}040$ bytes for \HQC{1}, \HQC{3}, and \HQC{5} due to the static buffer that keeps the transforms resident. Since these transforms were otherwise built on the stack at every call, the peak stack requirement instead decreases by $10{,}824$, $21{,}660$, and $29{,}944$ bytes. That is, caching essentially moves the transforms from the stack to the static region, so the total of static RAM and peak stack is nearly unchanged for \HQC{1} and \HQC{3} and only slightly higher for \HQC{5}. Unlike the stack, however, which is occupied only during an operation and released afterward, the cached transforms stay resident in RAM at all times, so in exchange for the lower peak stack they keep occupying that much memory even between KEM operations. Caching is therefore a technique that pays a permanently resident transform cache to gain speed.

\begin{table}[h!]
\centering
\caption{Memory usage of HQC implementations on Cortex-M4, with and without caching (in bytes).}
\label{tab:eval-memory}
\small
\resizebox{\textwidth}{!}{
\begin{tabular}{clrrrrr}
\toprule
\multirow{2}{*}{Parameter} & \multirow{2}{*}{Implementation} & \multirow{2}{*}{\makecell{Flash \\ (code+data)}} & \multirow{2}{*}{\makecell{RAM \\ (data+bss)}} & \multicolumn{3}{c}{Stack} \\
\cmidrule(r){5-7}
& & & & Keypair & Encaps & Decaps \\
\midrule
\multirow{4}{*}{\HQC{1}}
 & \cite{TCHES:CCC26} & $268{,}764$ & $2{,}824$ & $42{,}544$ & $52{,}864$ & $59{,}664$ \\
 & \cite{jlk25} & $294{,}972$ & $17{,}320$ & $43{,}752$ & $49{,}944$ & $56{,}744$ \\
 & Ours & $328{,}708$ & $14{,}344$ & $45{,}280$ & $55{,}432$ & $62{,}232$ \\
 & Ours (caching) & $329{,}796$ & $25{,}136$ & $-$ & $36{,}848$ & $51{,}408$ \\
\midrule
\multirow{4}{*}{\HQC{3}}
 & \cite{TCHES:CCC26} & $305{,}652$ & $2{,}824$ & $84{,}512$ & $105{,}736$ & $119{,}352$ \\
 & \cite{jlk25} & $340{,}868$ & $37{,}504$ & $86{,}440$ & $99{,}440$ & $113{,}056$ \\
 & Ours & $376{,}036$ & $37{,}504$ & $86{,}432$ & $107{,}496$ & $121{,}112$ \\
 & Ours (caching) & $377{,}188$ & $59{,}208$ & $-$ & $70{,}248$ & $99{,}452$ \\
\midrule
\multirow{4}{*}{\HQC{5}}
 & \cite{TCHES:CCC26} & $415{,}556$ & $2{,}824$ & $111{,}704$ & $140{,}976$ & $162{,}760$ \\
 & \cite{jlk25} & $246{,}436$ & $2{,}824$ & $111{,}696$ & $132{,}864$ & $154{,}648$ \\
 & Ours & $409{,}348$ & $2{,}824$ & $111{,}688$ & $140{,}960$ & $162{,}744$ \\
 & Ours (caching) & $410{,}052$ & $42{,}864$ & $-$ & $93{,}768$ & $132{,}800$ \\
\bottomrule
\end{tabular}
}
\end{table} 

\section{Conclusion and Future Work}
\label{sec:conclusion}

In this paper, we present optimizations that accelerate HQC on the ARM Cortex-M4. We find a $34\%$ sparser FAFFT modulus for \HQC{1}, and trim the VMOV count of the bit-sliced butterflies by $37.6$--$48.1\%$ at an unchanged EOR count via dirty-aware register allocation and XOR reordering. Predicated writes replace the masked support expansion of fixed-weight sampling, lowering its per-word cost from $22$ to $6$ cycles while remaining constant-time. Together, these reduce all three KEM operations by $22.0$--$34.6\%$ over the state of the art, and the optional public-key caching cuts encapsulation and decapsulation by up to a further $32.7\%$ and $18.9\%$ under a reused key.

Our VMOV optimizations are not specific to HQC. Work on the linear layers of block ciphers, including~\cite{xiang2020linear}, minimizes only the XOR count, yet on small in-order cores every spill costs an explicit move that this metric does not capture. Our method applies as a post-processing step to any such XOR-count-minimized implementation, such as the AES MixColumns.

\paragraph{Limitations and future work.}
Optimizing a fixed XOR sequence can miss globally cheaper code, as a slightly longer sequence may admit fewer register transfers; jointly searching over the sequence and the allocation is a natural next step. The optional caching benefits only deployments that repeatedly decapsulate under a fixed key, and keeps its transform cache permanently resident in RAM.

\bibliographystyle{alpha}
\bibliography{abbrev3,crypto,biblio}


\end{document}